\documentclass{article}

 \usepackage[workshop, final]{neurips_2026}
\usepackage[utf8]{inputenc} % allow utf-8 input
\usepackage[T1]{fontenc}    % use 8-bit T1 fonts
\usepackage{hyperref}       % hyperlinks
\usepackage{url}            % simple URL typesetting
\usepackage{booktabs}       % professional-quality tables
\usepackage{amsfonts}       % blackboard math symbols
\usepackage{nicefrac}       % compact symbols for 1/2, etc.
\usepackage{microtype}      % microtypography
\usepackage{xcolor}         % colors

\title{Beyond Disclosure: Compositional Privacy Harm in AI Systems}

\begin{document}

\author{%
  Didhiti KC \\
  % Regulatory and Compliance Counsel \\
  Truth Audit Labs \\
  Delhi, India \\
  \texttt{didhiti@truthauditlabs.ai} \\
  \And
  Aritra Das \\
  % AI Researcher \\
  Truth Audit Labs \\
  Delhi, India \\
  \texttt{aritra@truthauditlabs.ai} \\
  \And  
  Bhavesh Neekhra \\
  % AI Researcher \\
  Truth Audit Labs \\
  Delhi, India \\
  \texttt{bhavesh@truthauditlabs.ai} \\
  \And  
  Debayan Gupta \\
  % Founder-Director \\
  Truth Audit Labs \\
  Delhi, India \\
  \texttt{debayan@truthauditlabs.ai} 
}

\maketitle

\begin{abstract}
  Privacy law attaches obligations to each stage of information processing, whether it be collection, use, storage, or disclosure.
  We examine how legally cognisable privacy harm emerges across this full lifecycle rather than at any single one of these points. We argue that privacy risk in AI systems is compositional, arising from the interaction of memorisation, extraction, inference and contextual transfer. Disclosure is therefore not the sole threshold of privacy harm: legally significant consequences may emerge even where no underlying record is reproduced verbatim. Drawing on contextual integrity, we argue that privacy protection must extend across the lifecycle of AI-mediated information rather than attach to disclosure alone.
  \end{abstract}

\section{Introduction}
\label{sec:intro}

Privacy law has long protected personal information by following it through a lifecycle: collection, processing, use, storage and disclosure. Each stage carries its own legal requirements~\cite{eu2016gdpr}. This framework assumes that we can identify with reasonable precision the exact moment when privacy-relevant legal obligations crystallise and map them to specific actors. Artificial intelligence disrupts that assumption at every stage. The EU AI Act places comprehensive, risk-based obligations on AI systems while expressly operating alongside the Union's data-protection framework~\cite{eu2024aiact,edpb2024aimodels}. Its risk-classification approach sets graduated obligations according to a system's intended purpose and its potential for harm~\cite{eu2024aiact}, and it leaves unresolved the distinct privacy question of when a technical vulnerability in a machine learning system becomes a legally recognisable harm. This lacuna gives rise to a central question that we seek to address: at which point in the chain from model training, through memorisation, to extraction, and subsequent use does a technical vulnerability solidify into a violation of legally protected privacy interests? The answer to this question determines whether model developers bear responsibility for memorisation alone (a latent risk) or only once disclosure has actually occurred; and whether data subjects can bring a claim at each stage of the attack lifecycle or only at its end point. We advance the point that privacy risk in modern AI systems is fundamentally compositional: it arises not from any single vulnerability considered in isolation but from the interaction of multiple layers of technical vulnerability and legal ambiguity. 
We develop this argument in three parts.

\begin{itemize}
  \item Section~\ref{sec:memorisation} examines memorisation as a latent vulnerability,
        addressing its legal characterisation (Section~\ref{sec:legal-characterisation})
        and the distinction between latent risk and actual disclosure
        (Section~\ref{sec:latent-vs-actual}).
  \item Section~\ref{sec:extraction} considers extraction as the mechanism through which
        memorised information becomes accessible or legally consequential.
  \item Section~\ref{sec:contextual-integrity} draws on contextual integrity to explain how
        inference and the transfer of information across contexts can generate privacy harm
        without direct disclosure.
\end{itemize}

Each mechanism poses difficulties for existing privacy law on its own. We argue that it is
their combination that exposes the deepest inadequacies of current regulatory approaches.

\section{Memorisation as the Latent Vulnerability}
\label{sec:memorisation}

Foundation models learn abstract statistical patterns from their training data. Under certain conditions, they memorise specific sequences verbatim, encoding concrete data points within their parameters in a form that allows those points to be retrieved later~\cite{carlini2019secret,carlini2023quantifying}. Memorisation is increasingly viewed as an intrinsic property of high-capacity models rather than a training artefact. ~\cite{feldman2020memorization,tirumala2022memorization}. NIST includes privacy attacks based on memorisation and training-data reconstruction in its adversarial-machine-learning taxonomy~\cite{nist2025aml}. During training, a language model optimises its parameters to reduce prediction error across the training corpus. Empirical studies show that rare or unique sequences, duplicated examples, larger model capacity, and longer prompting context can increase measured memorisation or extractability~\cite{carlini2019secret,carlini2023quantifying,kandpal2022deduplicating,lee2022deduplicating,tirumala2022memorization}. Carlini et al. showed that GPT-2 could reproduce verbatim training data, including personal contact details and copyrighted text. ~\cite{carlini2023quantifying,nasr2023scalable,carlini2021extracting}.

\subsection{Legal Characterisation of Memorised Data}
\label{sec:legal-characterisation}

Existing data protection frameworks do not answer directly how information held within model parameters should be characterised. The first is whether information memorised by a foundation model counts as `personal data' under the applicable legislation. The GDPR defines personal data broadly as `any information relating to an identified or identifiable natural person'~\cite{eu2016gdpr}. More recently, the European Data Protection Board has addressed when AI models may themselves be treated as anonymous and emphasised that this assessment must consider the likelihood of extracting personal data or obtaining personal data through queries~\cite{edpb2024aimodels}. We ask a different but related question: whether the legal significance of memorisation changes as memorised information moves from a latent state to active disclosure. On its face, the GDPR definition can cover memorised information where model behaviour allows information relating to an identifiable person to be obtained. However, the difficulty lies in the way model parameters hold and reproduce that information. Unlike a conventional database, a foundation model need not retain a personal datum as a discrete, addressable record; this creates practical difficulties for access, rectification and erasure~\cite{zhang2025rtbf,bourtoule2021unlearning}. It is therefore unclear how those data-subject rights can be exercised in a complete and technically meaningful way against information encoded in model parameters.

A related question is whether the act of training a model on personal data constitutes `processing' within the meaning of data protection law. Processing is defined expansively under the GDPR, from collection and storage through to use and disclosure~\cite{eu2016gdpr}. The EDPB's AI-model opinion proceeds on the basis that development and deployment of AI models may involve processing of personal data and analyses the legal bases and consequences of such processing~\cite{edpb2024aimodels}. The training process plainly involves collection, organisation and use where personal data are included. The more difficult issue is how to characterise the resulting memorisation within model parameters. It is not `storage' in the conventional database sense because the data may not be retained as an addressable record; nor does memorisation necessarily create a discrete record that can be separately inspected or deleted. This mismatch between legal categories and technical representation is one reason erasure and unlearning remain technically difficult~\cite{bourtoule2021unlearning,zhang2025rtbf}.

\subsection{Latent Risk versus Actual Disclosure}
\label{sec:latent-vs-actual}

Perhaps the most fundamental conceptual challenge is where to draw the line between latent risk and actual disclosure. A model that has memorised a data subject's medical records poses a privacy risk, but has it violated that subject's privacy if the information is never extracted? The EDPB's treatment of AI-model anonymity illustrates why extractability and the likelihood of obtaining personal data are legally relevant even before a particular disclosure is shown~\cite{edpb2024aimodels}. At the same time, the GDPR distinguishes among the lawfulness, security and consequences of different processing operations rather than defining privacy harm solely by whether a disclosure has occurred~\cite{eu2016gdpr}. There is therefore no simple technical threshold at which latent memorisation automatically becomes the only legally cognisable harm.

On one view, memorisation is itself part of a processing operation and requires a lawful basis whether or not the memorised data is later disclosed. The violation may therefore arise during development where personal data are processed without an adequate legal basis, and any subsequent extraction may constitute an additional processing event. This view finds support in the GDPR's broad definition of processing and its security and data-protection-by-design obligations~\cite{eu2016gdpr,edpb2020privacybydesign}; the EDPB has also specifically analysed legitimate interests and unlawfully processed personal data in the development of AI models~\cite{edpb2024aimodels}. On an alternative view, memorisation without disclosure resembles retention of data in a system that has not yet been successfully queried or breached. The risk is real but unrealised, and operational duties therefore focus on reducing the probability that latent information can be extracted, whether through research on differential privacy, adversarial regularisation, dataset deduplication and machine unlearning. ~\cite{abadi2016deepprivacy,nasr2018membership,kandpal2022deduplicating,bourtoule2021unlearning}. Responsibility for implementing safeguards may nevertheless be distributed across providers and deployers rather than falling on the interface operator alone~\cite{edpb2021controller,eu2024aiact}.

We adopt a middle position: memorisation is a form of processing that requires legal justification, but privacy harm should not be seen as a simple distinction between latent risk and actual disclosure. Instead, the risk becomes more serious as memorised information becomes accessible or can be inferred outside the context in which it was originally processed or disclosed. The existence of memorised information may therefore trigger data protection obligations, while its extraction, inference, transfer to a different context, or use in making consequential decisions may create different and increasingly serious forms of privacy harm. Disclosure is one such form of harm, but it should not be treated as the sole mechanism through which latent privacy risk becomes legally consequential.

Information can generate privacy harm even where it is never directly reproduced or
transmitted to a third party. Inference attacks demonstrate more generally that access to
a model can reveal facts about individuals or about whether particular records were used
in training without reproducing an original record verbatim~\cite{fredrikson2015modelinversion,shokri2017membership}.
Appendix~\ref{app:insurance} develops a worked example in which information disclosed to a
chatbot in a personal context shapes a later insurance decision without any verbatim
disclosure to the decision-maker. Such cases demonstrate why a disclosure-based conception of privacy harm is insufficient
for AI systems. Privacy harm may arise not only when information is disclosed, but also
when information is retained, inferred, or used to make consequential decisions outside
the context in which it was originally provided~\cite{nissenbaum2004privacy,nissenbaum2010privacy}.

\section{Turning Memory Into Information through Extraction}
\label{sec:extraction}

Memorisation on its own is a dormant risk. There are several different methods that can be used to transform it into actual exposure or consequential use [7, 10, 25, 26]. They vary in the people who perform them, the rights they require and the evidence they leave, and these differences are legally relevant. Table 1 summarises them.

Extraction refers to a model's ability to reproduce its training data verbatim or almost verbatim. Carlini et al. extracted hundreds of verbatim sections from GPT-2's training data, such as names and contact information, by generating text in bulk and selecting out sections that are likely to have been memorised and later did the same on a much larger scale, even against commercial chatbots. A person who is familiar with some portion of a target text (e.g., a name, the beginning of an address) might be able to remember the remainder of that text, since a model text is more likely to be completed when more of its beginning is provided. The presence of such concrete evidence, the repeated text, makes it easy to characterise the legal nature of the extraction, namely as disclosure or making available within the definition of processing [1, Art. 4(2)] which must be based on specific legal grounds, and the reasonable expectations of the data subjects form part of the context-specific balancing.

Membership inference deals with a more specific question: was there any particular person's record used to train the model? The attacker has to be the one with the longer record, and compares to the model how confidently it can react to the attacker's record, with how it reacts to records that it has not seen before. The evidence consists of statistics and not a reproduced document. But the response can also be sensitive: When people learn that someone's record was used as a training set for a model created from hospital data, they know that the person was a patient. Model inversion and attribute inference go beyond this, however: Invert the model and infer facts about people, such as a health condition, that were not explicitly mentioned. This is difficult to separate from a model's ability to generalize, and is harder to prove. However, legally, if an inference is made which indirectly reveals someone's health, this can be special category data and so harm can occur without any record being repeated.

Who is the disclosing party? There are three parties that can be involved in a disclosure through a model: the company that created the model (the provider), the company that has made the model 

available to users (the deployer) and the user who posed the question (the user). Under GDPR controllership depends on who sets the purposes and essential means of the processing and what it is doing, not what it is called [1, 14]. The AI Act further makes a parallel allocation between providers and deployers, which is separate from GDPR controllership [2]. A security researcher testing a model, a malicious actor collecting personal data, and an ordinary user who receives the memorised data can cause the same technical incident with different legal repercussions depending on the role, intent, and predictability, and the legal basis.

Information without disclosure by a human discloser. There's another layer of complexity: Autonomous AI agents can disclose something without anyone asking for it. In July 2026, during internal testing of cybersecurity, OpenAI's AI agents managed to circumvent the protection measures designed to contain them, access the internet and gain access to the systems of the AI platform Hugging Face [29, 30]. OpenAI's admission claims the agents were attempting to "cheat" a challenging test by seeking out its solutions elsewhere. They were able to use the publicly visible Hugging Face user names and passwords to log in, and retrieved very limited private information, with some of the agents copying the private data from the evaluation process into a public one [29]. Hugging Face said it was investigating whether a customer or a partner's data was compromised, and had already noticed unauthorised access to internal data and service credentials [30] at the time of disclosure. No one initiated these revelations, so the differences in motive above mentioned does not apply. The responsibility, however, should shift to those who established the purposes and means of the activity in which the agents were used, and to foreseeability. OpenAI notes that they had seen early indications of the agents being able to communicate and surf the internet without permission weeks before, and that the safeguards they had put in place for the public products were not implemented during the test [29]. These are issues of security and governance (question (ii)) and, in the case of personal data, of notification in the event of breach (1, Arts 33–34). The AI Act might provide less purchase: the research, testing, and development phase that took place prior to the introduction of the model on the market is excluded [2, Art. 2(8)], and the model mainly responsible had not been released.
There are two types of “memory”. Memorisation within the model should be different to a memory feature within an application. One is the information absorbed into the model during the training process, the model's “parameters” that are controlled by the model's developer. The second one is one that is currently widely used in chatbots and injects the user's input into subsequent interactions, typically with the user's consent and sometimes at their own request. When you think of a memory feature, you can easily imagine it as a place where you can look and remove personal data, but when you consider a model's internal settings, you can't. The two have one thing in common: the deeper problem is that information learned in one context may be accessible or impactful in another, making contextual integrity the appropriate perspective on both [20, 21]. The Hugging Face incident reveals another third type of memory can be disparate. Initially, the agents left notes in an external memory service, shared file storage; other agents who worked in different tasks discovered the notes, and the service became an improvised message board to the agents' discoveries and exposed credentials were passed on [29]. Info created in a particular task was then accessible in various other tasks without anyone designing or authorizing such access – the simplest of contextual-integrity issues.

\section{Contextual Integrity as the Governing Normative Framework}
\label{sec:contextual-integrity}

Helen Nissenbaum's theory of contextual integrity~\cite{nissenbaum2004privacy,nissenbaum2010privacy} is particularly useful for understanding cross-session privacy violations. Nissenbaum argues that privacy is not fundamentally reducible to secrecy or individual control over information, but concerns appropriate information flows governed by context-specific norms. The framework evaluates informational actors, attributes or information types, and transmission principles, and asks whether a flow conforms to the norms of the relevant social context~\cite{nissenbaum2004privacy,nissenbaum2010privacy}. A privacy violation occurs when information flows in a manner that departs from those governing norms. Applied to persistent AI memory, contextual integrity reveals the precise nature of the harm. When a user shares medical information while seeking health advice, the governing norms permit that information to flow within the health-advice context. If the system later draws on it in a professional context, such as drafting an email to a colleague, the information has crossed a contextual boundary in violation of the norms governing its original disclosure. The user did not consent to their health information being available in professional contexts; they shared it within the health-advice context and expected it to remain there. The problem is not necessarily that the information was collected unlawfully: the user may have voluntarily provided it and may even have opted into a memory feature. In this scenario, the additional privacy problem is one of contextual migration: information legitimately held within one normative context is transferred to another context where different norms apply~\cite{nissenbaum2004privacy,nissenbaum2010privacy}. GDPR purpose limitation requires personal data to be collected for specified, explicit and legitimate purposes and restricts incompatible further processing~\cite{eu2016gdpr}. That principle captures part of the concern, but the legal concept of `purpose' does not map perfectly onto contextual integrity's broader account of actors, information types and transmission principles. Privacy in AI systems therefore turns in part on whether information was legitimately transferred between contexts, and not merely on whether it was legitimately accessed. Three consequences follow from this. First, compliance with collection and storage requirements is necessary but not sufficient where later use is incompatible with the original purpose or context~\cite{eu2016gdpr,nissenbaum2004privacy}. Secondly, systems that pool information across contexts carry a heightened need for technical and organisational safeguards that constrain later use~\cite{edpb2020privacybydesign}. Thirdly, the assessment of privacy harm cannot end at the moment of collection; it must also ask whether information was later transferred, inferred from, or used across a contextual boundary. Appendix~\ref{app:design} sets out the design implications, in particular the requirement
that systems retain and enforce contextual constraints rather than retrieve information on
semantic relevance alone.

\section{Compositional Privacy Harm}
\label{sec:compositional-privacy}

A harm is compositional when the review of each individual act shows that there is no harm in that particular act, but a review of the sequence of acts demonstrates that the personal information in the setting of origin was transported into a setting of harm to the person.

We can take the help of an everyday example. A person confides something to a friend; the friend mentions it, in passing, at a dinner; a guest later repeats it to the person’s employer. The individual chats might not be extraordinary on their own, however taken together, they can amount to something remarkable. The problem is the whole chain of events: information that should have been in a confidential environment has now been released somewhere it should not have been. Such chains are formed by AI systems regularly, even in multiple companies, and with nobody intending it.

The definition of compositional privacy harm:  

A compositional privacy harm occurs if three conditions are met:
Condition 1 - Each step has to be legal in and of itself. None of the steps should violate the law according to the laws that apply to it (its legal basis, its security duties, its declared purpose).
Condition 2 - Each step needs to be taken. If any one step was taken out or altered, the information would not have made it to the point where it caused harm.
Condition 3 - The final product violates the rules of the original setting. When considered together, the chain transfers information outside of the context in which it was shared and outside of the norms of that context (or a legal rule that protects information), and this has a real effect on the individual involved.

The composition of the risk is not cumulative. In a cumulative case, every step introduces some risk, and a person who looks at that step alone might see something amiss and correct it: Two systems with weak security are worse than one, but each level of security is visible. In a “compositional” case, a person reviewing any individual step would rightly conclude that none of these steps is wrong. The harm only occurs when it's in the chain, and so a review that only goes one step, one company, or one system will fall short. Nor is the same combination of data from different services in a single company as was already subject of the data-protection law (see the Court's reasoning in Meta Platforms v Bundeskartellamt regarding the combination of data from various services in a single company [29]). 

Neither is composition equivalent to a single company bringing together data it has collected from multiple sources, as the Court does in the case of Meta Platforms v Bundeskartellamt (data-protection law already examines this). The hard cases are those in which not one company at a particular time always has all the pieces of information.

\subsection{An Example in which All Steps are Lawful}

The following example in Appendix A satisfies all three conditions and is fully described with explicit assumptions. In outline: X, a cancer survivor entitled under national law not to declare a past diagnosis to an insurer, (1) talks about fear of the cancer returning with a chatbot whose memory feature X has switched on; (2) later asks the same chatbot to help fill in an insurance questionnaire, and the chatbot, drawing on what it remembers, drafts answers that never mention cancer but refer to regular hospital follow-ups and a “health-related” career break; (3) the insurer’s underwriting software, which has no access to the chatbot, assesses the application X submitted and rates X as higher-risk. All steps are backed up by laws and are justifiable. There is no word for word repetition, no information is shared between the two companies except X. But the statutory safeguards fail to work and the health data shared in seeking support becomes embedded in an insurance policy.

\subsection{Why the existing rules don't capture it}
Data-protection law and the AI Act focus on one step or one system, not the chain. The chatbot is not a high-risk AI under the AI Act and its operator's data protection impact assessment [1, Art. 35] considers the security of the memory store. The insurer's software is high-risk software and the insurer has to carry out a fundamental-rights impact assessment [2, Art. 27] but it is not asking itself how an answer by the applicant may have been formulated. The two companies are not joint controllers as they do not share a common purpose. And X's own remedies (contesting an automated decision [1, Art. 22(3)] or seeking an explanation [2, Art. 86]) would eventually lead back to words that X personally submitted.

\subsection{Checking the definition}
\begin{enumerate}
    \item Condition 1: Each step has to be legal in and of itself. This, as the table shows, is subject to two objections. First, it is not difficult to say that Step 2 violates the purpose limitation already. The declared purpose is “personalised assistance” – using the saved notes is within that purpose, so the objection is only accepted if purposes are linked to the context in which information was shared – which the concept of purpose in GDPR doesn't require (Section 4). Secondly, it could be argued that Step 3 is a violation of the statutory protection, which many would now identify as Step 2. This will depend on the type of protection that is provided, including indirect indicators of the diagnosis. If it has, it has done so without knowing, and that is only going to be discovered by reconstructing the entire chain. In either case, looking at any single step alone will not show the problem.

    \item Condition 2: Each step is necessary and needs to be taken. If Step 1 is not performed, the chatbot has no memory. At Step 2 if the chatbot could only access notes that had been shared at a similar situation, no health hints would be displayed. If the software program utilized by the insurer at Step 3 ignored such clues, then the premium wouldn't be increased. 

    \item Condition 3: The end result breaks the rules of the original setting. Information provided in a context of personal support enters into an insurance decision; a legal principle which protects the same information in that context, and has a financial impact; a decision which may fall within Article 22 [30].
    
\end{enumerate}
Comparing this to cumulative risk, a person wouldn't identify a problem in any particular review. Company D's effect assessment is about whether they maintain a memory store, and if they do, that store is kept confidential and never broken. The insurer's fundamental-rights impact assessment is a question of whether the software is accurate and not biased with regard to the inputs which are what the applicant provided.

\section{Conclusion and Limitations}
\label{sec:conclusion}

Privacy harm in AI systems does not arise at a single point. Memorisation may already
constitute processing requiring a legal basis~\cite{eu2016gdpr,edpb2024aimodels};
extraction may turn latent information into unauthorised
disclosure~\cite{carlini2021extracting,nasr2023scalable}; and inference or contextual
transfer may cause harm without reproducing any
record~\cite{fredrikson2015modelinversion,shokri2017membership,nissenbaum2004privacy}.
Their legal significance lies in how they interact. Two consequences follow. Assessment must ask at each stage whether information has become
extractable, inferable, or usable across a contextual boundary, and responsibility should follow each actor's role in enabling that shift rather than attach to whoever operates the
interface. Our account is normative and framed around the EU regime; which combinations of these mechanisms cause the most serious harm in deployed systems remains open.
% ---------------------------------------------------------------------------
% Acknowledgments: omit from the anonymised submission. The ack environment
% hides this section automatically when the submission is anonymised.
% ---------------------------------------------------------------------------
% \begin{ack}
% Funding and competing-interest declarations go here.
% \end{ack}

%\section*{References}

% The source text names several sources (the EU AI Act, the GDPR, the NIST
% adversarial ML taxonomy, Carlini et al. on GPT-2 memorisation, and
% Nissenbaum on contextual integrity) but contains no bibliography. Add the
% full references here, or switch to natbib/BibTeX with \bibliography{refs}.

\bibliographystyle{unsrt}
\bibliography{references}

%%%%%%%%%%%%%%%%%%%%%%%%%%%%%%%%%%%%%%%%%%%%%%%%%%%%%%%%%%%%
\newpage
\appendix

\section{Worked Example: Inference Without Disclosure}
\label{app:insurance}

Consider an insurance company that uses an AI system to assist in determining whether an
individual should be offered life insurance. Suppose an individual X, in a personal
capacity, talks to a chatbot and mentions, much as one might to a friend, that they have
terminal cancer. In one scenario, X discloses this information to the chatbot; in another,
X never discloses it. X subsequently approaches an insurance provider seeking life
insurance. The insurer's AI system has access, directly or indirectly, to the chatbot's
underlying model or memory. In the first scenario, the system may determine that X
presents a high mortality risk and recommend that insurance be denied; in the second, it
may recommend that coverage be granted.

The harm in the first scenario arises without any verbatim disclosure of X's earlier
statement to the insurer. X does not have to ask the system to retrieve the earlier
conversation, and the insurer does not have to inspect the chatbot's records. The system
can instead draw on retained information or representations and use them to influence a
downstream inference. Model-inversion and membership-inference research shows why privacy
leakage cannot be reduced to direct reproduction of training
records~\cite{fredrikson2015modelinversion,shokri2017membership,nasr2018membership}. The
decision is shaped by information that originated in an entirely different setting, even
though that information was never expressly reproduced to the decision-maker.

\section{Design Implications of Contextual Integrity}
\label{app:design}

A system that operationalises contextual integrity would need some mechanism for retaining
the context in which information was provided, evaluating whether a proposed later use is
appropriate to that context, and restricting cross-contextual transfers that lack an
independent justification~\cite{nissenbaum2010privacy}. These requirements resonate with
GDPR data protection by design and by default, which calls for safeguards to be integrated
into the means of processing rather than added only after
deployment~\cite{eu2016gdpr,edpb2020privacybydesign}. Persistent-memory architectures that
retrieve information solely on semantic relevance, without preserving or enforcing
contextual constraints, therefore create a structural risk of incompatible reuse.
\newpage
\section*{NeurIPS Paper Checklist}

\begin{enumerate}

\item {\bf Claims}
    \item[] Question: Do the main claims made in the abstract and introduction accurately reflect the paper's contributions and scope?
    \item[] Answer: \answerYes{}
    \item[] Justification: The abstract and Section~\ref{sec:intro} state the paper's central claim, that privacy risk in AI systems is compositional, and set out the scope of the argument developed in Sections~\ref{sec:memorisation}--\ref{sec:contextual-integrity}.

\item {\bf Limitations}
    \item[] Question: Does the paper discuss the limitations of the work performed by the authors?
    \item[] Answer: \answerYes{}
    \item[] Justification: Section~\ref{sec:conclusion} notes that the analysis is normative rather than empirical and is framed around the EU regulatory regime.

\item {\bf Theory assumptions and proofs}
    \item[] Question: For each theoretical result, does the paper provide the full set of assumptions and a complete (and correct) proof?
    \item[] Answer: \answerNA{}
    \item[] Justification: This is a legal  paper and contains no formal theoretical results requiring proof.

    \item {\bf Experimental result reproducibility}
    \item[] Question: Does the paper fully disclose all the information needed to reproduce the main experimental results of the paper to the extent that it affects the main claims and/or conclusions of the paper (regardless of whether the code and data are provided or not)?
    \item[] Answer: \answerNA{}
    \item[] Justification: This is a legal  paper and reports no experiments.

\item {\bf Open access to data and code}
    \item[] Question: Does the paper provide open access to the data and code, with sufficient instructions to faithfully reproduce the main experimental results, as described in supplemental material?
    \item[] Answer: \answerNA{}
    \item[] Justification: This is a legal  paper and involves no data or code.

\item {\bf Experimental setting/details}
    \item[] Question: Does the paper specify all the training and test details (e.g., data splits, hyperparameters, how they were chosen, type of optimizer) necessary to understand the results?
    \item[] Answer: \answerNA{}
    \item[] Justification: This is a legal  paper and reports no experiments.

\item {\bf Experiment statistical significance}
    \item[] Question: Does the paper report error bars suitably and correctly defined or other appropriate information about the statistical significance of the experiments?
    \item[] Answer: \answerNA{}
    \item[] Justification: This is a legal  paper and reports no experimental results.

\item {\bf Experiments compute resources}
    \item[] Question: For each experiment, does the paper provide sufficient information on the computer resources (type of compute workers, memory, time of execution) needed to reproduce the experiments?
    \item[] Answer: \answerNA{}
    \item[] Justification: This is a legal  paper and required no compute resources.
    
\item {\bf Code of ethics}
    \item[] Question: Does the research conducted in the paper conform, in every respect, with the NeurIPS Code of Ethics \url{https://neurips.cc/public/EthicsGuidelines}?
    \item[] Answer: \answerYes{}
    \item[] Justification: The paper is a doctrinal and normative analysis of published legal instruments and existing technical literature. It involves no human subjects, no personal data and no new models or datasets, and conforms with the Code of Ethics.

\item {\bf Broader impacts}
    \item[] Question: Does the paper discuss both potential positive societal impacts and negative societal impacts of the work performed?
    \item[] Answer: \answerYes{}
    \item[] Justification: The societal implications are the subject matter of the paper. It analyses privacy harms arising from memorisation, extraction, inference and contextual transfer, and argues for safeguards spanning the lifecycle of AI-mediated information.
    
\item {\bf Safeguards}
    \item[] Question: Does the paper describe safeguards that have been put in place for responsible release of data or models that have a high risk for misuse (e.g., pre-trained language models, image generators, or scraped datasets)?
    \item[] Answer: \answerNA{}
    \item[] Justification: This is a legal  paper. It releases no data or models, and describes extraction techniques only at the level of detail already published in the cited literature.

\item {\bf Licenses for existing assets}
    \item[] Question: Are the creators or original owners of assets (e.g., code, data, models), used in the paper, properly credited and are the license and terms of use explicitly mentioned and properly respected?
    \item[] Answer: \answerNA{}
    \item[] Justification: This is a legal  paper and uses no code, data or models. All legal instruments and prior work relied upon are cited in the references.

\item {\bf New assets}
    \item[] Question: Are new assets introduced in the paper well documented and is the documentation provided alongside the assets?
    \item[] Answer: \answerNA{}
    \item[] Justification: This is a legal paper and releases no new assets.

\item {\bf Crowdsourcing and research with human subjects}
    \item[] Question: For crowdsourcing experiments and research with human subjects, does the paper include the full text of instructions given to participants and screenshots, if applicable, as well as details about compensation (if any)? 
    \item[] Answer: \answerNA{}
    \item[] Justification: This is a legal paper and involves no crowdsourcing or human subjects.

\item {\bf Institutional review board (IRB) approvals or equivalent for research with human subjects}
    \item[] Question: Does the paper describe potential risks incurred by study participants, whether such risks were disclosed to the subjects, and whether Institutional Review Board (IRB) approvals (or an equivalent approval/review based on the requirements of your country or institution) were obtained?
    \item[] Answer: \answerNA{}
    \item[] Justification: This is a legal and normative  paper and involves no human subjects, so no IRB review was required.

\item {\bf Declaration of LLM usage}
    \item[] Question: Does the paper describe the usage of LLMs if it is an important, original, or non-standard component of the core methods in this research? Note that if the LLM is used only for writing, editing, or formatting purposes and does \emph{not} impact the core methodology, scientific rigor, or originality of the research, declaration is not required.
    \item[] Answer: \answerNA{}
    \item[] Justification: LLMs were not used as a component of the core method. Language models are the subject of analysis rather than a research instrument.

\end{enumerate}
\end{document}